\documentclass[aps,prl,twocolumn,showpacs,groupedaddress]{revtex4}
\usepackage{graphics}
\usepackage{graphicx}

\begin{document}


\title{Violation of Energy Conservation in Boson and Fermion Fields\\on Subwavelength Nano-Scale}


\author{S.V. Kukhlevsky}
\affiliation{Department of Physics, University of P\'ecs,
Ifj\'us\'ag u.\ 6, H-7624 P\'ecs, Hungary}


\begin{abstract}
The Hamiltonians describing the energy nonconservation in boson and fermion multimode fields under quantum interference have been derived. We show that violation of the energy conservation is accompanied by the nonconservation of momentum, number of particles and field charge. The phenomena could be observed in Young's double-slit subwavelength (nanometer-scale) setup.
\end{abstract}


\maketitle

%
\maketitle
%
%
The energy conservation law, which is one of the most important
laws in physics, states that energy can be converted from one form
to another, but it cannot be created or destroyed. The energy
conservation is a mathematical consequence of the shift
symmetry of time. More abstractly, energy is a generator of a continuous time-shift
symmetry of the physical system under study. When a physical
system has a time-shift symmetry, Noether's theorem~\cite{Noeth} implies the
existence of a conserved current. The thing that "flows" in the
current is the energy, the energy is the generator of the
symmetry group~\cite{Noeth,Bye}.

The energy conservation affects all systems without exceptions,
for an example, classical~\cite{Nott,Land,Jack} and
quantum~\cite{Mont,Bere,Itz,Ryde,Grei,Wein,Pes,Scul} fields.
Recently, we have shown that the cross-correlation energy
associated with classical interference violates the energy
conservation in an ensemble of EM modes having different
phases~\cite{Kukh}. The ensemble does not obey
the shift symmetry of time (the energy nonconservation does not contradict
the energy conservation law) when the time shifts are different
for each of the modes. The energy nonconservation should not
be confused with the uncertainty principle. In quantum mechanics,
lack of commutation of the time derivative operator with the time
operator itself mathematically results in an uncertainty principle
for time and energy: the longer the period of time, the more
precisely energy can be defined. Although, quantum theory in general,
and the uncertainty principle specifically, do not violate energy
conservation, we have shown that the quantum interference could
create or destroy energy for infinite time. In the present paper,
the Hamiltonians describing the energy nonconservation in boson and
fermion multimode fields under quantum interference are derived.
We show that violation of the energy conservation is
accompanied by the nonconservation of momentum, number of
particles and field charge. The phenomena could be
observed in Young's two-source subwavelength setup.

With the objective of deriving the Hamiltonians that describe the
energy nonconservation in boson and fermion fields under quantum
interference, let us begin with the canonical quantization of a
Boson scalar field (spin $s$ = 0) described by the Lagrangian
density ${\cal
L}=({\partial}_{\mu}{\psi}^*)({\partial}^{\mu}{\psi})-m^2{\psi}^*{\psi}$~.
The dynamics of the field is determined by the Euler-Lagrange
equation of motion. Due to the space-time shift symmetry, the
fields $\psi$ and $\psi ^*$ obey the conservation laws for the
energy and
momentum~\cite{Noeth,Bye,Nott,Mont,Land,Jack,Bere,Itz,Ryde,Grei,Wein,Pes,Scul}.
The Lagrangian density is invariant under $U(1)$ local gauge
transformation ($\psi \rightarrow \psi '=e^{i\varphi} \psi $)
given rise to the conserved current. The continuity equation for
the 4-vector of the current density $j_\mu = m(\psi ^* (\partial
_\mu \psi) -(\partial _\mu \psi ^*)\psi  )$ yields the conservation
of the field charge $Q={\int}j_0d^3x$. The Hamiltonian of the
system is given by
\begin{eqnarray}
{\cal H}={\int} \left( \frac {\partial {\psi ^*}}{\partial t} \frac {\partial {\psi}}{\partial t} +
\nabla {\psi ^*}\cdot \nabla {\psi } + m^2{\psi}^*{\psi} \right)d^3x.
\end{eqnarray}
The secondary quantization of the field is performed by replacing the fields $\psi$ and $\psi ^*$
by the respective multimode field operators $\hat \psi$ and $\hat \psi ^{\dagger }$, where
\begin{eqnarray}
\hat \psi=\sum_{\mathbf{k}}{(2{\varepsilon}_{\mathbf{k}})^{-1/2}}({\hat a}_{\mathbf{k}}e^{-i{\mathbf{k}
{\mathbf{r}}}}+{\hat b}_{\mathbf{k}}^{\dagger}e^{i{\mathbf{k}{\mathbf{r}}}}),
\end{eqnarray}
\begin{eqnarray}
\hat \psi ^{\dagger }
=\sum_{\mathbf{k}}{(2{\varepsilon}_{\mathbf{k}})^{-1/2}}({\hat
a}_{\mathbf{k}}^{\dagger} e^{i{\mathbf{k}{\mathbf{r}}}}+{\hat
b}_{\mathbf{k}}e^{-i{\mathbf{k}{\mathbf{r}}}}),
\end{eqnarray}
and ${\varepsilon}_{\mathbf{k}}=({\mathbf{k}}^2+m^2)^{1/2}$. Here,
${\hat a^{\dagger}}_{\mathbf{k}}$, ${\hat a}_{\mathbf{k}}$ and
${\hat b^{\dagger}}_{\mathbf{k}}$, ${\hat b}_{\mathbf{k}}$ are
respectively creation and destruction operators for boson particles and
antiparticles inside a volume $V=1$. The operators satisfy the canonical
Bose commutation relations. Equation (1) yields the
conventional Hamiltonian:
\begin{eqnarray}
{\hat {\cal H}}=\sum_{\mathbf{k}}{\varepsilon}_{\mathbf{k}}(\mathbf{\hat {\cal N}}_{\mathbf{k}}+
\mathbf{\hat {\bar {\cal  N}}}_{\mathbf{k}}+1).
\end{eqnarray}
Here, $\mathbf{\hat {\cal N}}_{\mathbf{k}}={{\hat a}_{\mathbf{k}}^{\dagger }}{{\hat a}_{\mathbf{k}}}$
and $\mathbf{\hat {\bar {\cal  N}}}_{\mathbf{k}}={{\hat b}_{\mathbf{k}}^{\dagger }}{{\hat b}_{\mathbf{k}}}$
are the number operators for particles and antiparticles, respectively.

We now consider a multimode field $\psi=\sum_{\mathbf{k}}{\psi}_{\mathbf{k}}e^{i\varphi_{\mathbf{k}}}$,
namely a superposition of $N$ phase-coherent modes with different phases $\varphi _{\mathbf{k}}$.
The field does not obey the above-mentioned symmetries (conservation laws)~\cite{Noeth,Bye,Nott,Mont,Land,Jack,Bere,Itz,Ryde,Grei,Wein,Pes,Scul} if the shifts are different
for each of the modes. The secondary quantization is performed by replacing the fields $\psi$ and $\psi ^*$ in (1) by the operators
\begin{eqnarray}
\hat \psi=\sum_{n=1}^N{(2\varepsilon  )^{-1/2}}({\hat a}_{\mathbf{k}_n}e^{-i{\mathbf{k}_n
{\mathbf{r}}}-{i\varphi_{\mathbf{k}_n}}}+{\hat b}_{\mathbf{k}_n}^{\dagger}e^{i{\mathbf{k}_n{\mathbf{r}}}+
{i\varphi_{\mathbf{k}_n}}})
\end{eqnarray}
and
\begin{eqnarray}
\hat \psi ^{\dagger } =\sum_{n=1}^N{(2\varepsilon  )^{-1/2}}({\hat a}_{\mathbf{k}_n}^{\dagger}e^{i{\mathbf{k}_n
{\mathbf{r}}}+{i\varphi_{\mathbf{k}_n}}}+{\hat b}_{\mathbf{k}_n}
e^{-i{\mathbf{k}_n{\mathbf{r}}}-{i\varphi_{\mathbf{k}_n}}}).
\end{eqnarray}
Thus the Hamiltonian (1) for the multimode field, an ensemble of noninteracting particles
and antiparticles, has the form
\begin{eqnarray}
\hat {\cal H}=\sum_{n=1}^N\hat {\cal H}_{nn}+\sum_{n\neq m}^{N}\sum_{m\neq{n}}^{N}\hat {\cal H}_{nm},
\end{eqnarray}
where
\begin{eqnarray}
\hat {\cal H}_{nn}={\varepsilon}_{\mathbf{k}_n}(\mathbf{\hat {\cal N}}_{\mathbf{k}_n}+
\mathbf{\hat {\bar {\cal  N}}}_{\mathbf{k}_n}+1)
\end{eqnarray}
and
\begin{eqnarray}
\hat {\cal H}_{nm}={{\frac{{\hat a^{\dagger }}_{\mathbf {k_n} } {\hat
a}_{\mathbf {k_m}}}{( {\varepsilon}_n {\varepsilon}_m)^{-1/2}}}
{\int}e^{-i{\mathbf{k}_n{\mathbf{r}-i\varphi_{n}}}}e^{i{\mathbf{k}_m{\mathbf{r}+i\varphi_{m}}}}d^3x}\nonumber \\
+{{\frac{{\hat b}_{\mathbf {k_n} }{\hat b^{\dagger }}_{\mathbf {k_m}}}{( {\varepsilon}_n {\varepsilon}_m)^{-1/2}}}
{\int}e^{i{\mathbf{k}_n{\mathbf{r}+i\varphi_{n}}}}e^{-i{\mathbf{k}_m{\mathbf{r}-i\varphi_{m}}}}d^3x}.
\end{eqnarray}
The procedure of secondary quantization for momentum $P$,
number of particles ${\cal N}$ and field charge $Q$ is similar.
The first term in Eq.~7 corresponds to the conventional Hamiltonian
(4) of a Boson scalar
field~\cite{Mont,Land,Jack,Bere,Itz,Ryde,Grei,Wein,Pes,Scul}. The
second term, which is neglected in the field theory~\cite{Mont,Land,Jack,Bere,Itz,Ryde,Grei,Wein,Pes,Scul},
does associate with the quantum interference and the
positive or negative cross-correlation energy.
The interference phenomenon and phase
correlation between the undistinguishable particles of different
modes (${\mathbf{k}_n} \neq {\mathbf{k}_m}$) are taken into
account by using the nonconventional (Bose-type) commutation
relations $[{\hat a}_{{k}_n},{\hat a^{\dagger }}_{{k}_m}]$ =
$[{\hat b}_{{k}_n},{\hat b^{\dagger }}_{{k}_m}]$ = 1; the other
operator pairs commute. One can use also the equivalent
commutation relations: $[{\hat a}_{{k}_n},{\hat a^{\dagger
}}_{{k}_m}]$ = $[{\hat b}_{{k}_n},{\hat b^{\dagger }}_{{k}_m}] =
\pm {e^{-i\varphi_{m}+i\varphi_{n}}}$. The two descriptions are
different only in the vacuum energy. The first relations yield
nonzero fluctuations of energy about its zero ensemble average for
a vacuum state, $\langle {\cal E}\rangle {\neq} 0$ (see, (7-9)). In
the case of $[{\hat a}_{{k}_n},{\hat a^{\dagger }}_{{k}_m}]$ =
$[{\hat b}_{{k}_n},{\hat b^{\dagger }}_{{k}_m}] =
-{e^{-i\varphi_{m}+i\varphi_{n}}}$ and ${\mathbf {k}_n}={\mathbf
{k}_m}$,  the fluctuation of boson energy about its zero ensemble
average for a vacuum state is zero, $\langle {\cal E}\rangle=0$.

So far we have considered the Hamiltonian that takes into account
the quantum interference in a field of scalar ($s$ = 0) Boson
particles. The quantization of a field of Boson particles having
the spin $s > 0$ is similar to the above-considered case of scalar
particles. As an example, we derive the Hamiltonian of a Boson
vector (spin $s$ = 1) field of noninteracting particles that
corresponds to an EM field. An EM field is described by the
Lagrangian density ${\cal L}=(-1/4){{\psi}_{\mu \nu }
}{{\psi}^{\mu \nu } }$, where the 4-tensor $\psi ^{\mu \nu }$ is
the field tensor $F^{\mu \nu }$; the 4-vector $\psi ^\mu $ is the
4-potential $A^\mu
$~\cite{Nott,Mont,Land,Jack,Bere,Itz,Ryde,Grei,Wein,Pes,Scul}. The
operators of the field have the form:
\begin{eqnarray}
{\hat \psi}_\mu =\sum_{\mathbf {k}, \alpha } {(2\varepsilon  )^{-1/2}} ({\hat a}_{\mathbf{k}\alpha }{u_{\mu }^{\alpha
}}e^{-i{\mathbf{k} {\mathbf{r}}}}+{\hat
b}_{\mathbf{k}\alpha }^{\dagger}{u_{\mu }^{\alpha
^*}}e^{i{\mathbf{k}{\mathbf{r}}}})
\end{eqnarray}
and
\begin{eqnarray}
{\hat \psi ^{\dagger }}_\mu =\sum_{\mathbf {k}, \alpha
}{(2\varepsilon  )^{-1/2}} ({\hat a}_{\mathbf{k}\alpha }^{\dagger
}{u_{\mu }^{\alpha ^*}}e^{i{\mathbf{k} {\mathbf{r}}}}+{\hat
b}_{\mathbf{k}\alpha }{u_{\mu }^{\alpha
}}e^{-i{\mathbf{k}{\mathbf{r}}}}),
\end{eqnarray}
where ${u_{\mu }^{\alpha ^*}}$ is the unit 4-vector of
polarization; the index $\alpha $ corresponds to the two
independent polarizations. Owing to the identity of the photons
and anti-photons, ${\hat a}={\hat b}$ and ${\hat
a}^{\dagger}={\hat b}^{\dagger}$. The secondary quantization of
the multimode field, an ensemble of $N$ phase-coherent modes with
different phases $\varphi _{\mathbf{k}}$, yields the Hamiltonian
\begin{eqnarray}
\hat {\cal H}=\sum_{n=1,\alpha }^N\hat {\cal H}_{nn\alpha }+\sum_{n\neq m,
\alpha }^{N}\sum_{m\neq n, \alpha }^{N}\hat {\cal H}_{nm \alpha },
\end{eqnarray}
where
\begin{eqnarray}
\hat {\cal H}_{nn,\alpha }={\varepsilon}_{\mathbf{k}_n}(\mathbf{\hat {\cal N}}_{\mathbf{k}_n \alpha }+ \frac{1}{2})
\end{eqnarray}
and
\begin{eqnarray}
\hat {\cal H}_{nm,\alpha \alpha }={{\frac{{\hat a^{\dagger }}_{\mathbf {k_n \alpha } } {\hat
a}_{\mathbf {k_m \alpha }}}{2( {\varepsilon}_n {\varepsilon}_m)^{-1/2}}}
{\int}e^{-i{\mathbf{k}_n{\mathbf{r}-i\varphi_{n}}}}e^{i{\mathbf{k}_m{\mathbf{r}+i\varphi_{m}}}}d^3x}\nonumber \\
+{{\frac{{\hat a}_{\mathbf {k_n \alpha } }{\hat a^{\dagger }}_{\mathbf {k_m \alpha }}}
{2( {\varepsilon}_n {\varepsilon}_m)^{-1/2}}}
{\int}e^{i{\mathbf{k}_n{\mathbf{r}+i\varphi_{n}}}}e^{-i{\mathbf{k}_m{\mathbf{r}-i\varphi_{m}}}}d^3x }.
\end{eqnarray}
We assume the commutation relation $[{\hat a}_{{k}_n\alpha },{\hat a^{\dagger }}_{{k}_m\alpha }]=1$;
the other operator pairs commute.

The quantization of a multimode field of Fermion particles is similar to
the above-considered Bose particles. As an example, consider a spinor ($s$ = 1/2) field described
by the Dirac Lagrangian density ${\cal L}=i\bar \psi \gamma ^{\mu }\partial _{\mu}\psi -m\bar \psi \psi $.
It is not necessary to repeat all the quantization procedures. The secondary quantization is performed
by replacing the fields $\psi$ and $\bar \psi$ in the free Dirac
field~\cite{Mont,Land,Jack,Bere,Itz,Ryde,Grei,Wein,Pes} by the operators:
\begin{eqnarray}
{\hat \psi} =\sum_{\mathbf {k}, \sigma  } {(2\varepsilon  )^{-1/2}} ({\hat c}_{\mathbf{k}\sigma  }{u_{k \sigma
}}e^{-i{\mathbf{k} {\mathbf{r}}}}+{\hat
d}_{\mathbf{k}\sigma  }^{\dagger}{u_{-k -\sigma
}}e^{i{\mathbf{k}{\mathbf{r}}}})
\end{eqnarray}
and
\begin{eqnarray}
{\hat {\bar \psi }} =\sum_{\mathbf {k}, \sigma  } {(2\varepsilon  )^{-1/2}}
({\hat c}_{\mathbf{k}\sigma }^{\dagger }{{\bar u}_{k \sigma }}e^{i{\mathbf{k} {\mathbf{r}}}}+{\hat
d}_{\mathbf{k}\sigma }{{\bar u}_{-k -\sigma }}e^{-i{\mathbf{k}{\mathbf{r}}}}),
\end{eqnarray}
where the summation is performed over all values $\bf{k}$ and $\sigma =\pm 1/2$.
The secondary quantization of the multimode field, an ensemble of $N$ phase-coherent modes
having different phases $\varphi _{\mathbf{k}}$, yields the Hamiltonian
\begin{eqnarray}
\hat {\cal H}=\sum_{n=1,\sigma  }^N\hat {\cal H}_{nn\sigma  }+
\sum_{n\neq m, \sigma  }^{N}\sum_{m\neq n, \sigma  }^{N}\hat {\cal
H}_{nm \sigma},
\end{eqnarray}
where
\begin{eqnarray}
\hat {\cal H}_{nn,\sigma  }={\varepsilon}_{\mathbf{k}_n}(\mathbf{\hat {\cal N}}_{\mathbf{k}_n \sigma  }
+\mathbf{\hat {\bar {\cal  N}}}_{\mathbf{k}_n\sigma  } -1)
\end{eqnarray}
and
\begin{eqnarray}
\hat {\cal H}_{nm,\sigma \sigma}={{\frac{{\hat c^{\dagger }}_{\mathbf {k_n \sigma  } } {\hat
c}_{\mathbf {k_m \sigma  }}}{( {\varepsilon}_n {\varepsilon}_m)^{-1/2}}}
{\int}e^{-i{\mathbf{k}_n{\mathbf{r}-i\varphi_{n}}}}e^{i{\mathbf{k}_m{\mathbf{r}+i\varphi_{m}}}}d^3x}\nonumber \\
-{{\frac{{\hat d}_{\mathbf {k_n \sigma  } }{\hat d^{\dagger }}_{\mathbf {k_m \sigma  }}}
{( {\varepsilon}_n {\varepsilon}_m)^{-1/2}}}
{\int}e^{i{\mathbf{k}_n{\mathbf{r}+i\varphi_{n}}}}e^{-i{\mathbf{k}_m{\mathbf{r}-i\varphi_{m}}}}d^3x }.
\end{eqnarray}
The interference phenomenon and phase correlation between the undistinguishable fermions of
different modes (${\mathbf{k}_n} \neq {\mathbf{k}_m}$) are taken
into account by using the nonconventional (Fermi-type)
anticommutation rules $\{{\hat c}_{{k}_n \sigma }, {\hat
c^{\dagger }}_{{k}_m \sigma }\}$ = $\{{\hat d}_{{k}_n\sigma
},{\hat d^{\dagger }}_{{k}_m\sigma }\}$ = 1; the other operator
pairs are anticommutative. In the case of $\{{\hat a}_{{k}_n},{\hat a^{\dagger }}_{{k}_m}\}$ =
$\{{\hat b}_{{k}_n},{\hat b^{\dagger }}_{{k}_m}\} =
-{e^{-i\varphi_{m}+i\varphi_{n}}}$ and ${\mathbf
{k}_n}={\mathbf {k}_m}$, the fluctuation of fermion energy about its zero ensemble
average for a vacuum state is zero.

The cross-correlation integrals (9), (14) and (19) could be
interpreted as exchange ones. The integrals describe the
quantum exchange interference associated with the
indistinguishability of identical particles. The quantum exchange
of particles is somewhat similar to the exchange of virtual
particles for a short time ($\Delta t \leq 1/\Delta \varepsilon $) 
in perturbation theory. The cross-correlation integrals
have nonzero values if $\mathbf {k}_n \approx \mathbf {k}_m$ or
$V=$ $\Delta x\Delta y\Delta z \leq 1 / (k_n-k_m)_x
(k_n-k_m)_y (k_n-k_m)_z$. In agreement with principle of the
indistinguishability of individual bosons, the commutation
relations introduced above have the canonical form if $\mathbf
{k}_n = \mathbf {k}_m$. In the case of fermions, one should take
into account the Pauli exclusion principle.

It is worth to note that all properties of an ensemble of bosons
described by the first term in (7) are undistinguishable from the
conventional Hamiltonian (4). The first term $\hat {\cal H}=\hat {\cal
H}_{\mathbf {k}_1}\otimes \hat {\cal H}_{\mathbf {k}_2}... \otimes
\hat {\cal H}_{\mathbf {k}_N}$is responsible for the quantum
interference of the modes with themselves. The term does not take
into account the interference between the different modes. The
quantum state $\phi  \in \hat {\cal H}$ is a pure state because it
is separable, $\phi = \phi_{\mathbf {k}_1}\otimes  \phi_{\mathbf
{k}_2}...\otimes  \phi_{\mathbf {k}_N}$. The energy $\langle
{\cal E}\rangle={\sum_{n=1}^N}{\varepsilon}_{\mathbf{k}_n}(\langle
\mathbf{{\cal N}}_{\mathbf{k}_n}\rangle+ \langle \mathbf{{\bar
{\cal  N}}}_{\mathbf{k}_n}\rangle +1)$, momentum $\langle
{\cal P} \rangle={\sum_{n=1}^N}{\mathbf{k}_n}(\langle
\mathbf{{\cal N}}_{\mathbf{k}_n}\rangle+ \langle \mathbf{{\bar
{\cal  N}}}_{\mathbf{k}_n}\rangle +1)$, a total number
$\langle
{\cal N} \rangle={\sum_{n=1}^N}(\langle \mathbf{{\cal
N}}_{\mathbf{k}_n}\rangle+\langle \mathbf{{\bar {\cal
N}}}_{\mathbf{k}_n}\rangle +1)$ of bosons, and field charge
$\langle Q \rangle={\sum_{n=1}^N}(\langle \mathbf{{\cal
N}}_{\mathbf{k}_n}\rangle-\langle \mathbf{{\bar {\cal
N}}}_{\mathbf{k}_n}\rangle -1)$ are conserved under interfering
("mixing") the quantum states $\phi_{\mathbf {k}_n}$.

The behavior of the field described by the full Hamiltonian (7) is different from the usual Hamiltonian (4).
The field quantum state $\phi$ is an entangled state because
$\phi \neq \phi_{\mathbf {k}_1}\otimes  \phi_{\mathbf {k}_2}...\otimes  \phi_{\mathbf {k}_N}$.
The Hamiltonian (7) takes into account the phase correlation between the modes under the intermode interference.
The positive or negative cross-correlation energy associated with the second term in (7) violates
energy conservation in the field. Indeed, the energy can be created or destroyed in an ensemble of
coherent modes by changing (shifting) the phases $\varphi_{\mathbf {k}_n}$ if the shifts are different for each of the modes. The phase modification under the
free-space propagation or reflection of the mode does not require additional energy. The interference
of modes completely destroys the energy if the modes interfere destructively in all points of
a physical system. The interference creates energy if the modes interfere only constructively.
For instance, in the case of ${\mathbf{k}_n} = {\mathbf{k}_m}$ (the commutation relations have the canonical form), the total energy depends on the values
$\varphi_n$ and $\varphi_m$: $0\leq{\langle {\cal E} \rangle}\leq {\varepsilon}_{\mathbf{k}}N^2
(\langle \mathbf{{\cal N}}_{\mathbf{k}}\rangle+\langle \mathbf{{\bar {\cal  N}}}_{\mathbf{k}}\rangle +1)$
(see, (5-9)). The total number of particles and antiparticles can vary from zero to
$N^2(\langle \mathbf{{\cal N}}_{\mathbf{k}}\rangle+\langle \mathbf{{\bar {\cal  N}}}_{\mathbf{k}}\rangle +1)$.
In the case of $\varphi_n-\varphi_m=\pi$ for the mode pairs, the interference completely destroys
the energy (${\langle {\cal E} \rangle}$=0), momentum (${\langle {\cal P} \rangle}$=0),
charge ($\langle Q \rangle=0$) and the probability to find
bosons (${\langle \phi \mid  \hat \psi ^{\dagger }\hat \psi \mid \phi\rangle}=0$); here, we have discarded
irrelevant constants associated with the vacuum field. The field has zero energy in both the
vacuum and non-vacuum states under the commutation relations
$\{{\hat a}_{{k}_n},{\hat a^{\dagger }}_{{k}_m}\}$ = $\{{\hat b}_{{k}_n},{\hat b^{\dagger }}_{{k}_m}\}$
= $-{e^{-i\varphi_{m}+i\varphi_{n}}}$.
The interference creates the maximum of the energy
($\langle
{\cal E}\rangle={\varepsilon}_{\mathbf{k}}N^2(\langle \mathbf{{\cal N}}_{\mathbf{k}}\rangle+
\langle \mathbf{{\bar {\cal  N}}}_{\mathbf{k}}\rangle +1$)), momentum
($\langle
{\cal P} \rangle={\mathbf{k}}N^2(\langle \mathbf{{\cal N}}_{\mathbf{k}}\rangle+
\langle \mathbf{{\bar {\cal  N}}}_{\mathbf{k}}\rangle +1$)), a number of particles
($\langle {\cal N} \rangle=N^2(\langle \mathbf{{\cal N}}_{\mathbf{k}}\rangle+\langle
\mathbf{{\bar {\cal  N}}}_{\mathbf{k}}\rangle +1$)) and
charge ($\langle Q \rangle=N^2(\langle \mathbf{{\cal N}}_{\mathbf{k}}\rangle -
\langle \mathbf{{\bar {\cal  N}}}_{\mathbf{k}}\rangle -1$))
if $\varphi_n-\varphi_m=0$.
In the case of the phase-noncorrelated modes, the energy $\langle {\cal E}
\rangle={\varepsilon}_{\mathbf{k}}N(\langle \mathbf{{\cal N}}_{\mathbf{k}}\rangle+
\langle \mathbf{{\bar {\cal  N}}}_{\mathbf{k}}\rangle +1)$, momentum $\langle {\cal P}
\rangle={\mathbf{k}}N(\langle \mathbf{{\cal N}}_{\mathbf{k}}\rangle+
\langle \mathbf{{\bar {\cal  N}}}_{\mathbf{k}}\rangle +1)$, a number of
bosons $\langle {\cal N} \rangle=N(\langle \mathbf{{\cal N}}_{\mathbf{k}}\rangle+\langle
\mathbf{{\bar {\cal  N}}}_{\mathbf{k}}\rangle +1)$,
and field charge $\langle Q \rangle=
N(\langle \mathbf{{\cal N}}_{\mathbf{k}_n}\rangle-\langle \mathbf{{\bar {\cal  N}}}_{\mathbf{k}_n}\rangle -1)$
are conserved under the interference (see, (4-9)). If the phase-coherent modes have a spatial dependence
appropriate for a cavity resonator, the model predicts the same result; the resonator modes are
orthogonal to each other.

The nonconservation of energy, momentum, number of particles and
field charge in the multimode boson field could be observed in
Young's double-slit subwavelength setup. Let us consider the
two-mode field; each of the phase-coherent modes contains one
boson. The field describes a bi-boson (two entangled bosons). In
conventional Young's setup, the two plane waves (modes) generated
by the pinholes separated by the distance ${\Lambda}>>\lambda$
have different wave vectors, ${\mathbf k_1}\neq {\mathbf k_2}$
\cite{Kukh}. The cross-correlation integrals (9) vanish if
${\mathbf k_1}\neq {\mathbf k_2}$. Correspondingly, the energy
$\langle {\cal
E}\rangle={\sum_{n=1}^2}{\varepsilon}_{\mathbf{k}_n} (\langle
\mathbf{{\cal N}}_{\mathbf{k}_n}\rangle+\langle \mathbf{{\bar
{\cal  N}}}_{\mathbf{k}_n}\rangle +1)$, momentum $\langle {\cal P}
\rangle={\sum_{n=1}^2}{\mathbf{k}_n}(\langle \mathbf{{\cal
N}}_{\mathbf{k}_n}\rangle+ \langle \mathbf{{\bar {\cal
N}}}_{\mathbf{k}_n}\rangle +1)$, a total number $\langle {\cal N}
\rangle={\sum_{n=1}^2}(\langle \mathbf{{\cal
N}}_{\mathbf{k}_n}\rangle+ \langle \mathbf{{\bar {\cal
N}}}_{\mathbf{k}_n}\rangle +1)$ of bosons, and field charge
$\langle Q\rangle={\sum_{n=1}^2} (\langle \mathbf{{\cal
N}}_{\mathbf{k}_n}\rangle-\langle \mathbf{{\bar {\cal
N}}}_{\mathbf{k}_n}\rangle -1)$ are conserved; here $\langle
\mathbf{{\cal N}}_{\mathbf{k}_n}\rangle= \langle \mathbf{{\bar
{\cal  N}}}_{\mathbf{k}_n}\rangle =1$. In the case of Young's
subwavelength (${\Lambda}<<\lambda$, ${\mathbf k_1}$ =
${\mathbf k_2}$) system, the interference completely destroys the
energy (${\langle {\cal E} \rangle}$=0), momentum (${\langle {\cal
P} \rangle}$=0), charge ($\langle Q \rangle=0$) and the
probability to find bosons (${\langle \phi \mid  \hat \psi
^{\dagger }\hat \psi \mid \phi\rangle}=0$) if
$\varphi_1-\varphi_2=\pi$. If
$\varphi_1-\varphi_2=0$, the interference creates the maximum of
the energy ($\langle {\cal
E}\rangle={\varepsilon}_{\mathbf{k}}4(\langle \mathbf{{\cal
N}}_{\mathbf{k}}\rangle+\langle \mathbf{{\bar {\cal
N}}}_{\mathbf{k}}\rangle +1$)), momentum ($\langle {\cal P}
\rangle={\mathbf{k}}4(\langle \mathbf{{\cal
N}}_{\mathbf{k}}\rangle+\langle \mathbf{{\bar {\cal
N}}}_{\mathbf{k}}\rangle +1$)), charge ($\langle
Q\rangle=4(\langle \mathbf{{\cal N}}_{\mathbf{k}}\rangle - \langle
\mathbf{{\bar {\cal  N}}}_{\mathbf{k}}\rangle -1$)) and a number
of particles ($\langle {\cal N} \rangle=4(\langle \mathbf{{\cal
N}}_{\mathbf{k}}\rangle+\langle \mathbf{{\bar {\cal
N}}}_{\mathbf{k}}\rangle +1$)); here $\langle \mathbf{{\cal
N}}_{\mathbf{k}_n}\rangle=\langle \mathbf{{\bar {\cal
N}}}_{\mathbf{k}_n}\rangle =1$. The result is different from the
generally accepted opinion that the energy of two bosons, for
instance photons, is always constant. According to
our model, the energy of two noncorrelated bosons only is
preserved, $\langle {\cal
E}\rangle={\sum_{n=1}^2}{\varepsilon}_{\mathbf{k}_n}(\langle
\mathbf{{\cal N}}_{\mathbf{k}_n}\rangle+\langle \mathbf{{\bar
{\cal  N}}}_{\mathbf{k}_n}\rangle +1)$. This is in agreement with
the fact that the frequency (energy) of a biphoton depends on the
experimental configuration (the values of ${\mathbf k_n}$ and
$\varphi_n$) of the biphoton generation~\cite{Sou}. The
interference of a single boson with itself does not create or
destroy energy. In such a case, one of the modes characterized by
${\mathbf k_1}$ or ${\mathbf k_2}$ does not contain the boson.
Notice that the phase conditions required for the creation or
destruction of photon energy in Young's two-source
experiment~\cite{Scul,Baha} can be easily realized experimentally by
using two subwavelength fibres with different refraction indexes.
The energy nonconservation in a multimode photon field is relevant
also to a Dicke quantum model \cite{Dick} of superradiance
emission of a subwavelength ensemble of atoms. In the model, the
wave vectors of the light waves produced by the atoms in the
far-field zone are practically the same, $\mathbf
k_n\approx\mathbf k$. In addition to the  superradiance, our model
predicts the total destruction of energy in the Dicke ensemble of
atoms at the condition $\varphi_n-\varphi_m=\pi$ for the atom
(photon) pairs. Finally, our model shows that a quantum entangled
state of photons is preserved on passage through a subwavelength
aperture array; the values ${\mathbf k}$ and $\varphi_k$ do not
change (see, (12-14)). The propagation of an entangled state
through a lens, however, destroys the
entanglement by the well-known modification of the wave vectors
${\mathbf k}$ and phases $\varphi_{\mathbf k}$. Such a behavior is
in agreement with the experiment \cite{Alte}.

It should be stressed that our results do not contradict the
conservation laws attributed to the space-time shift symmetry and
the $U(1)$ local gauge symmetry. The superposition of
phase-coherent modes
$\psi=\sum_{\mathbf{k}}{\psi}_{\mathbf{k}}e^{i\varphi_{\mathbf{k}}}$
with different phases $\varphi _{\mathbf{k}}$ and the respective
field operators do not obey the symmetries (conservation laws)
if the shifts are different for each of the modes.
The quantum state of the field is an entangled state. The
interference phenomenon and phase correlation between the
undistinguishable particles of different modes (${\mathbf{k}_n}
\neq {\mathbf{k}_m}$) were taken into account by using the
nonconventional commutation relations $[{\hat a}_{{k}_n},{\hat
a^{\dagger }}_{{k}_m}]$ = $[{\hat b}_{{k}_n},{\hat b^{\dagger
}}_{{k}_m}]$ = 1. In the case $\mathbf {k}_n = \mathbf {k}_m$, the
commutation relations have the canonical form. It worth noting
that $[{\hat a}_{{k}_n},{\hat a^{\dagger }}_{{k}_m}]$ $\approx$
$[{\hat b}_{{k}_n},{\hat b^{\dagger }}_{{k}_m}]$ $\approx 0$ for
very large values of $\langle{{\cal N}}_{\mathbf{k}}\rangle$ and
$\langle{\bar {\cal N}}_{\mathbf{k}}\rangle$. In this case, the
quantum energy calculated by (7) is the same as the "classical"
value (1) according to the correspondence principle, which states
that the quantum and "classical" treatments must agree for a very
large number of particles. We have presented formulas for the
coherent modes. One can easily rewrite Eqs.~(1-19) for the
incoherent or partially coherent modes. The positive or negative
cross-correlation energy and the corresponding positive or
negative~\cite{Feyn} probability can be easily demonstrated for
the boson and fermion fields described by the Hamiltonians (12)
and (17). Notice that the energy, momentum, number of particles
and electrical charge do not conserve in the fermion multimode
field under the interference. In the case of electrons, the phase
conditions required for the creation or destruction of energy in
Young's two-source experiment can be easily realized
by placing a solenoid between the slits like that
in the Bohm-Aharonov setup~\cite{Aha}.

This study was supported by the Hungarian Scientific
Research Foundation (OTKA, Contract No T046811).

\end{document}